\documentclass[aps,superscriptaddress,reprint,prl]{revtex4-1}
\usepackage{epsfig,graphics,amssymb,amsmath,subeqnarray,mathrsfs,color,xcolor}
\usepackage{color,epsfig,graphics,amssymb,amsmath,subeqnarray,graphicx,amsthm,subfigure,mathrsfs}
\usepackage[colorlinks=true, citecolor=red, linkcolor=blue ]{hyperref}
\def\S{{\bf S}}
\def\E{{\bf E}}
\def\e{{\bf e}}

\def\d{{\rm d}}\def\p{{\bf p}}
\def\i{{\rm i}}

\def\x{{\bf x}}
\def\n{{\bf n}}
\def\f{{\bf f}}
\def\u{{\bf u}}
\def\I{{\bf I}}
\def\in{{\int\!\!\!\int_{\partial V}}}
\def\bsigma{\boldsymbol{\sigma}}
\def\bSigma{{\boldsymbol \Sigma}}

\newcommand{\pard}[2]{\frac{\partial #1}{\partial #2}}
\newcommand{\change}[1]{{\color{black} #1}}
\begin{document}
\title{A reciprocal method to compute the stresslet of self-propelled bodies}
\title{Stresslets induced by active swimmers}
\author{Eric Lauga} 
\email{e.lauga@damtp.cam.ac.uk}     
\affiliation{Department of Applied Mathematics and Theoretical Physics, University of Cambridge, CB3 0WA, United Kingdom.}
\author{S\'ebastien Michelin}
\email{sebastien.michelin@ladhyx.polytechnique.fr}
\affiliation{LadHyX -- D\'epartement de M\'ecanique, Ecole Polytechnique -- CNRS, 91128 Palaiseau, France.}
\date{September 9, 2016}
 
\begin{abstract}
Active  particles disturb the fluid around them as force dipoles, or stresslets, which govern their collective dynamics.  Unlike  swimming speeds, the stresslets of active particles are rarely determined due to the lack of a suitable theoretical framework \change{for arbitrary geometry}. We propose  a \change{general} method, based  on the reciprocal theorem of Stokes flows, to  compute stresslets as integrals of the velocities on the  particle's surface, which we illustrate  for spheroidal chemically-active particles. Our method will allow tuning the stresslet of artificial swimmers and tailoring their collective motion in complex environments.

\end{abstract}
\maketitle

The study of swimming microorganisms could be hailed as the biophysics `poster child' due to the  ability  of classical  physics  to provide robust quantitative predictions  \cite{lighthill76,lp09}. Mathematical theories developed from first principles have \change{been able to} quantitatively capture the locomotion of bacteria \cite{lauga16}, spermatozoa \cite{fauci06}, algae \cite{stocker} as well as their collective dynamics \cite{koch} and their interactions with complex chemical environments  \cite{bergbook}. In addition, self-propelling cells and artificial active particles \cite{paxton06,golestanian07} have provided  the soft matter community with model systems to  discover new physics \cite{ramaswamy10,marchetti13}.

The primary quantity of interest for a  swimming body, and what most theory work  focuses on,  is its swimming speed, $U$.  A wealth of experimental  data exists for a large variety of biological cells  \cite{brennen77}. Mathematical methods have been developed to predict  swimming speeds, in particular resistive-force \cite{cox70} and slender-body theory \cite{johnson80}. These solve for the force distribution along an organism by taking  advantage of the linearity of the Stokes equations for the  fluid  flow to determine  the swimming kinematics without requiring a full computation of the  flow. With its swimming speed known, a swimmer is  then seen to display long-time effective diffusion at a rate $D\sim U^2\tau$
where the time scale $\tau$ is the relevant one for loss of orientation, be it thermal noise or  cell  tumbling \cite{Berg1993}.

Beyond the   swimming speed, an equally important characteristic of a self-propelled body is  its stresslet. Since  cells and active particles  swim without applying net forces  to the surrounding fluid, the flows they induce  have the symmetry of a  force dipole and decay spatially as $\sim 1/r^2$. Formally,  the velocity field in the laboratory frame at a location $\x$ away from a swimmer can generically be written in the far field as   $\u=-3(\x\cdot\S\cdot\x)\x /8\pi\mu r^5$, where $r=|\x|$ and $\S$ is the  trace-free second rank  stresslet tensor which is symmetric  when the swimmer does not apply any net moment \cite{Batchelor1970}. For axisymmetric swimming along a direction $\e$, then one obtains $\S=S(\e\e-\tfrac{1}{3}\I)$, and the sign of $S$ allows to distinguish 
between two types of swimmers: 
pusher cells with $S<0$ are pushed from behind and include most flagellated bacteria; in contrast, puller cells with  $S>0$ are pulled forward, e.g.~the biflagellated algae {\it Chlamydomonas}.

 The stresslets of self-propelling cells and active particles have been the subject of much less attention than their swimming speeds, but they are no less important. The magnitudes and signs of stresslets govern pattern formation and interactions  in populations of cells  \cite{guell88}, dictate which type of  swimmer suspension is unstable and displays nonlinear fluctuations   \cite{saintillan13}, and the physics of collective locomotion 
 \cite{dombrowski04,sokolov07}. The   stresslet also controls the interactions of active organisms   with their  environment   \cite{berke08,Drescher_etal2009}, enhanced transport through biological fluids  \cite{wu00,jepson13} and the rheology of active fluids \cite{saintillan_shear}.

If the stresslet of active swimmers  is so important, why do so few studies attempt to determine its value? The difficulty lies in the fact that, unlike the swimming speed which is purely a kinematic quantity, the stresslet  includes information about both kinematics and dynamics as it is formally given by an integral on the surface of the swimmer  of both  instantaneous surface velocities and surface stresses \cite{Batchelor1970}. Solving for  both velocities and surface stresses can be done numerically using the boundary element method \cite{ishikawa07_bacteria}, but typically not analytically. An alternative method consists in measuring, or computing, the flow far from the swimmer and fitting it to the expected
 stresslet, but so far this has been done only with the bacterium {\it E.~coli} \cite{drescher11} and requires an experimental apparatus able to distinguish the far field flow   from measurement noise.

% Phrase ˆ ajouter:
% The Lorentz Reciprocal Theorem has since then been applied to a large variety of hydrodynamic problems~\cite{leal_review}, to take into account Marangoni, inertial or viscoelastic effects. 
In this paper, we propose a theoretical method to compute the stresslet induced by active swimmers. Twenty years ago, Stone \& Samuel derived an integral theorem to determine the swimming speed of any swimmer using an auxiliary problem of  rigid-body motion  \cite{stone96}. \change{This result relies on the Lorentz Reciprocal Theorem which has proved popular in the hydrodynamics community to compute Marangoni, inertial or viscoelastic effects on the motion of particles, drops and bubbles~\cite{leal_review1980,raja2010,pak2014b}, and even the flux of boundary-driven channel flows~\cite{michelin2015c}.} We show that a similar approach may be undertaken to determine the value of the stresslet for active particles \change{of arbitrary shape}. We   derive a new integral theorem, involving an auxiliary problem of a passive rigid particle in a linear flow, allowing the determination of the full stresslet tensor. After \change{validating it for the classical} problems of swimming of a sphere (squirming) and locomotion of an active rod, we show that the theorem allows to determine exactly, for the first time, the stresslet induced by  ellipsoidal swimmers of any aspect ratio. We apply our results to  phoretic particles and discover how the pusher-puller transition  depends on the geometry of the particle.

In  seminal work, Batchelor~\cite{Batchelor1970}   showed that the contribution of an active particle of surface $\partial V$ to the bulk stress, i.e. the so-called stresslet tensor $\S$, is given by 
\begin{eqnarray}\label{stresslet2}
S_{ij}&=&\in\left[\frac{1}{2}(
x_j \sigma_{ik}n_k 
+ x_i  \sigma_{jk}n_k )
\right.\\
&&\left.
\quad-\frac{1}{3}(x_k\sigma_{kl}n_l)\delta_{ij}
-\mu (u_i n_j + u_j n_i)
\right] \d A.\nonumber
\end{eqnarray}
For active particles or cells prescribing a relative surface velocity $\u^s$ (or swimming gait), the second part of this integral  can be directly evaluated (its value  does not depend on the swimming velocity).  In contrast, the first part involves the surface traction, $\bsigma\cdot\n$, which  in general can only be obtained by solving for the flow  everywhere.  In order to calculate   this first part of the stresslet integral,  we use the reciprocal theorem of Stokes flow written as \cite{leal}
\begin{equation}\label{rec}
\in  u_i\bar \sigma_{ij}n_j\d A =\in  \bar u_i\sigma_{ij}n_j\d A,
\end{equation}
where we choose the dual flow field $(\bar\u, \bar\bsigma)$, a  solution of Stokes' equations that decays at infinity, to satisfy $\bar \u = \E \cdot \x$ on the particle's boundary where  $\E$ is a constant, symmetric and traceless second-order tensor, \change{and the origin of $\x$ is chosen so that the particle is force- and torque-free}.  The solution $(\bar\u,\bar\bsigma)$ is thus the instantaneous perturbation flow induced by the presence of the same active particle when stationary  in a linear flow field, i.e.~$\u=-\E\cdot\x+\bar\u$.
The associated stress field can be formally written as $\bar\bsigma (\x)\equiv \mu \bSigma(\x) : \E$ where $\bSigma$ is a dimensionless 4th-order  tensor  symmetric with respect to the first two and last two indices  (due to the symmetries of $\bar\bsigma$ and $\E$). 

After  changing  indices, the left-hand side of Eq.~\eqref{rec} becomes
\begin{equation}\label{left}
\in  u_i\bar \sigma_{ij}n_j\d A
= 
\mu \left(\in  n_l u^s_k   \Sigma_{klij}\d A\right)E_{ij},
\end{equation}
whereas the right-hand side is
\begin{equation}\label{right}
\in  \bar u_i\sigma_{ij}n_j\d A   
 = \left(\in \frac{1}{2}\left(x_j\sigma_{ik}n_k+x_i\sigma_{jk}n_k\right) \d A\right) E_{ij},
 \end{equation}
where the term in parenthesis has been replaced by its symmetric part since $\E$ is symmetric. 
Equating Eqs.~\eqref{left} and~\eqref{right}, for {any} trace-free symmetric tensor $\E$, we obtain
\begin{equation}
\in \frac{1}{2}\left(x_j\sigma_{ik}n_k+x_i\sigma_{jk}n_k\right) \d A
 =
 \mu \in  n_l u^s_k   \Sigma_{klij}\d A,
 \end{equation}
up to an isotropic second-order tensor. %, $C  \delta_{ij}$.  
The trace-free portion of this result is given by 
\begin{eqnarray}
&&\in\left[\frac{1}{2}(
x_j \sigma_{ik}n_k 
+ x_i  \sigma_{jk}n_k )
-\frac{1}{3}(x_k\sigma_{kl}n_l)\delta_{ij}
\right] \d A\nonumber \\
&&= \mu \in n_l u^s_k  \left( \Sigma_{klij} -  \frac{1}{3}\Sigma_{klmm}\delta_{ij}\right)\d A.\label{eq:tractionpart}
\end{eqnarray}
%Note that in order for Eq.~\eqref{eq:tractionpart} to be -- similarly to Eq.~\eqref{stresslet2} --  independent of the reference frame, the origin $\x=0$ in the linear flow must be chosen so that the total force on the particle is zero.

Combining Eqs~\eqref{stresslet2} and \eqref{eq:tractionpart}, we finally obtain the stresslet tensor $\S$ as 
\begin{equation}\label{final}
\frac{S_{ij}}{\mu} =  \in n_l u^s_k  \left( \Sigma_{klij} -  \frac{\delta_{ij}}{3}\Sigma_{klmm}
-\delta_{ik}\delta_{jl} - \delta_{il}\delta_{jk}
\right)
\d A.
\end{equation}
The  result in Eq.~\eqref{final} is an explicit integral of the prescribed, or measured, surface velocity $\u^s$, \change{ and does not depend on the swimming velocity of the particle -- similarly to Eq.~\eqref{stresslet2}.} Provided $\bSigma$ can be computed once and for all for the same geometry (either analytically or numerically), this results allows one to directly compute the stresslet generated by the active particle or cell {for any surface velocity} and without actually solving the associated flow problem. 

This integral  formulation can first be used to recover classical results, starting with the stresslet induced by a squirming sphere~\citep{blake1971}.  The dual flow field, $\bar\u$, for a sphere \change{of radius $a$} in a linear flow is a classical solution given by~\cite{leal}
\begin{align}
\bar\u& = a^5\frac{\E\cdot \x}{r^5} + 
\frac{5(\x\cdot\E\cdot\x)\x}{2}\left(\frac{a^3}{r^5}-\frac{a^5}{r^7}\right),
\\
\bar{p} & = 5a^3\mu \frac{\x\cdot\E\cdot\x}{r^5},
\end{align}
From this, the  tensor $\bar\bsigma$ and thus $\bSigma$ may be easily evaluated~\cite{SI}. Using Eq.~\eqref{final}, the stresslet is  obtained  as
\begin{equation}
\mathbf{S}=\mu \in  \left(-\frac{5}{2}(n_j\delta_{ik}+n_i\delta_{jk})+\delta_{ij}n_k\right)u^s_k\d A.
\end{equation}
For an axisymmetric squirming sphere~\citep{blake1971}, the prescribed slip velocity is purely tangential $\u^s=u^s(\zeta)\mathbf{e}_\theta$  ($\zeta=\cos\theta$ in spherical polar coordinates). In that case, the stresslet simplifies to
\begin{align}\label{eq:stresslet_sphere}
\mathbf{S}&=-\frac{5\mu}{2}\in \left(\n\u^s+\u^s\n\right)\d A
\end{align}
and finally
\begin{align}\label{eq:stresslet_sphere2}
\mathbf{S}&=15\pi\mu a^2\left(\e_z\e_z-\frac{1}{3}\I\right)\int_{-1}^1 u^s(\zeta)\zeta\sqrt{1-\zeta^2}\d \zeta.
\end{align}
This result is  equivalent to decomposing the slip velocity onto the canonical squirming modes, with the second mode providing the intensity of the stresslet~\cite{blake1971,michelin2014,pak2014}.

Another classical model is the active rod. A  rod of length $L$ and unit direction vector $\p$ imposes an axisymmetric slip velocity $\u^s=\alpha(s)\p$ in its reference frame, with $-L/2\leq s\leq L/2$   the arc-length measured along the rod. To determine the stresslet, the force distribution acting on a rigid rod   in a linear flow $\u = -\E\cdot \x$ must be computed. The  integral to calculate in Eq.~\eqref{final} is 
\begin{equation}
\mu \in n_l u_k   \Sigma_{klij}
\d A = 
\int _L u_k \int_{\partial V_R}\mu n_l    \Sigma_{lkij}\d A,
\end{equation}
where $n_l    \Sigma_{lkij}$ is obtained through the 
 force per unit length acting on the rigid rod as 
 \begin{equation}
\bar f_k = \left(\int_{\partial V_R}\mu n_l \Sigma_{klij}\d A \right)E_{ij}.
\end{equation}
 The force density, $\bar \f $, can be obtained using resistive-force theory \cite{cox70,lp09}  (with $\x=s\p$)
 \begin{equation}
\bar \f (s,t)= s\zeta_\perp\left(\frac{\p\p}{2}-\I\right)\cdot\E\cdot\p,
\end{equation}
%Eric: ok with Eq 15
and thus
\begin{equation}
\int_{\partial V_R}\mu n_l \Sigma_{klij} \d A= s\zeta_\perp \left(\frac{p_ip_k}{2}-\delta_{ik}\right)p_j,
\end{equation}
where $\zeta_\perp$ is the perpendicular drag coefficient for the rod \cite{cox70,lp09}. 
Using these results, Eq.~\eqref{final} becomes finally
\begin{equation}\label{rod_recip}
\S
=
-\left(\frac{1}{2}\zeta_\perp U_0   
\int_Ls \alpha(s) \d s\right) \left(\p\p-\frac{1}{3}\I\right),
 \end{equation}
which is identical to the result of a direct calculation~\cite{SI}.

The power of the integral method in Eq.~\eqref{final} may be demonstrated on  problems where a direct calculation of $\S$ is \change{not tractable analytically}. Motivated by recent work on phoretic swimmers, we illustrate this for an axisymmetric active spheroidal particle (or swimmer) of axis $\e_z$ and semi-axes $a$ and $b$. In this case, the flow field can still be computed as a superposition of \change{spheroidal harmonics}~\cite{kanevsky2010}, but a direct calculation of the tensor $\S$ \change{from a projection of $\u^s$ on the relevant harmonics is much more difficult.} In contrast, the integral formulation  allows to determine $\S$ exactly \change{and explicitly, for an arbitrary $\u^s$}.

Focusing on an axisymmetric distribution of slip velocity at the boundary, the stresslet $\mathbf{S}$ is a trace-less symmetric tensor invariant by rotation around $\e_z$ and must therefore be of the form
$ \mathbf{S}=S\left(\e_z\e_z-\frac{1}{3}\I\right)$. 
It  is thus sufficient to use as dual velocity field  the axisymmetric solution of Stokes' equations  decaying at infinity and satisfying $\bar\u=E\left(\e_z\e_z-\frac{1}{3}\I\right)\cdot\x$ on the spheroid's boundary with arbitrary $E$. 
Following classical work~\cite{jeffery1922}, the dual velocity field $\bar\u$ and associated fluid force on the particle $\bar\bsigma\cdot\n$ can be found explicitly. In particular we have
\begin{equation}
\bar\bsigma\cdot\n=2\mu\left[\frac{2EG(\xi)}{9F(\xi)}\I +\left(1-\frac{2}{3F(\xi)}\right){\mathbf{E}}\right]\cdot\n,
\end{equation}
where $\xi\equiv a/b$  is the  aspect ratio and the function $F$ is
\begin{equation}
F(\xi)=\frac{1}{(\xi^2-1)^2}\left[-3\xi^2+\frac{\xi(1+2\xi^2)}{\sqrt{1-\xi^2}}\cos^{-1}\xi\right],
\end{equation}
while the function $G$ is not required for what follows \cite{SI}. Using our  integral formulations,  one then easily  obtains
\begin{eqnarray}\label{eq:stresslet_ell}
\mathbf{S}&=&-\frac{2\mu}{3F(\xi)}\in\left(\u^s\n+\n\u^s\right)\d A,
\end{eqnarray}
with $\u^s$ the prescribed slip velocity at the particle's boundary. This new result  is valid for both prolate ($\xi\geq 1$) and oblate ($\xi\leq 1$) spheroids (note that $F(1)=4/15$, agreeing with Eq.~\ref{eq:stresslet_sphere}).  

We use spheroidal polar coordinates $(\tau,\zeta,\phi)$ with  $(x,y)=k\sqrt{\tau^2\mp 1}\sqrt{1-\zeta^2}(\cos\phi,\sin\phi)$ (for prolate and oblate spheroids, respectively), 
$z=k\zeta\tau$, $k=\sqrt{\mathscr{S}|\xi^2-1|/2\pi H(\xi)}$ with $\mathscr{S}$, the surface area of the spheroid, and
\begin{equation}
H(\xi)=1+\frac{\xi^2}{\sqrt{\xi^2-1}}\cos^{-1}\left(\frac{1}{\xi}\right).
\end{equation}
The surface of the particle  is then defined by $\tau=\tau_0=\xi/\sqrt{|\xi^2-1|}$. For an active particle that prescribes an axisymmetric slip velocity $\u^s=u^s(\zeta)\e_\zeta$,  the strength of the stresslet is then obtained as the integral
\begin{align}\label{stresslet_ell2}
S=-\frac{2\mathscr{S}\mu}{F(\xi)H(\xi)}\int_{-1}^1 u^s(\zeta)\zeta\sqrt{\frac{\xi^2(1-\zeta^2)}{\zeta^2+\xi^2(1-\zeta^2)}}\d \zeta.
\end{align}

We can now apply  this result to an autophoretic spheroidal particle  releasing a solute  of diffusivity $D$ with  fixed flux $\mathcal{A}(\zeta)$ along  its boundary. Interactions between the  particle surface  and the solute leads to a phoretic fluid  slip velocity, $\u=\mathcal{M}(\zeta)(I-\n\n)\cdot\nabla C$, induced along its boundary \cite{Anderson1989}.  When solute advection  is negligible, its concentration, \change{$C$}, is solution to the diffusive problem 
\begin{equation}\label{eq:cproblem}
\change{D\nabla^2 C=0,\quad D\e_\tau\cdot\left.\nabla C\right|_{\partial V}=-\mathcal{A}(\zeta),\quad C(\infty)=0.}
\end{equation}
With the new integral result above, Eq.~\eqref{stresslet_ell2}, we can now obtain the stresslet generated by the catalytic particle {without solving the actual Stokes flow problem}. Since Laplace's equation is separable in spheroidal coordinates,  
Eq.~\eqref{eq:cproblem} can  be solved explicitly for  $c$ as
\begin{eqnarray}
\change{C(\tau,\zeta)}&=&-\sum_{n=0}^\infty \frac{k(2n+1)C_n(\tau)}{2DC_n'(\tau_0)} {\cal I}_n(\xi)L_n(\zeta)\label{eq:c},\\
{\cal I}_n(\xi)&=&\int_{-1}^1\mathcal{A}(\zeta)\sqrt{\zeta^2+\xi^2(1-\zeta^2)}L_n(\zeta)\d\zeta,
\end{eqnarray}
where $C_n(\tau)=Q_n(\tau)$ or $Q_n(\i\tau)$ for prolate and oblate spheroids, respectively, and \change{$L_n$ and $Q_n$ are the Legendre polynomials and function of the second kind, respectively}.  The general expression for the resulting stresslet of a spheroid, Eq.~\eqref{stresslet_ell2}, can now be evaluated as $\S=S\left(\e_z\e_z-\frac{1}{3}\I\right)$, with strength
\begin{equation}
S=-\frac{\mu\xi}{DF(\xi)}\sqrt{\frac{8\pi\mathscr{S}}{H(\xi)}}\int_{-1}^1 \frac{\mathcal{M}(\zeta) \zeta(1-\zeta^2)}{\zeta^2+\xi^2(1-\zeta^2)}\pard{c}{\zeta}\d \zeta.
\end{equation}
Using Eq.~\eqref{eq:c}, the stresslet intensity $S$ of a catalytic spheroidal particle of aspect ratio $\xi$ is finally obtained as
\begin{equation}\label{finalfinal}
S=\frac{\mu\mathscr{S}}{D}\frac{\xi\sqrt{|\xi^2-1|}}{F(\xi)H(\xi)}\sum_{n=1}^\infty {\cal I}_n(\xi){\cal J}_n(\xi){\cal K}_n(\xi),
\end{equation}
with
\begin{align}
{\cal J}_n(\xi)&=\int_{-1}^1 \frac{\zeta (1-\zeta^2)M(\zeta)L_n'(\zeta)}{\zeta^2+\xi^2(1-\zeta^2)} \d \zeta,\\
{\cal K}_n(\xi)&=\frac{(2n+1)C_n\left(\frac{\xi}{\sqrt{|\xi^2-1|}}\right)}{C_n'\left(\frac{\xi}{\sqrt{|\xi^2-1|}}\right)}\cdot
\end{align}

\begin{figure*}[t]
\begin{center}
\includegraphics[width=.99\textwidth]{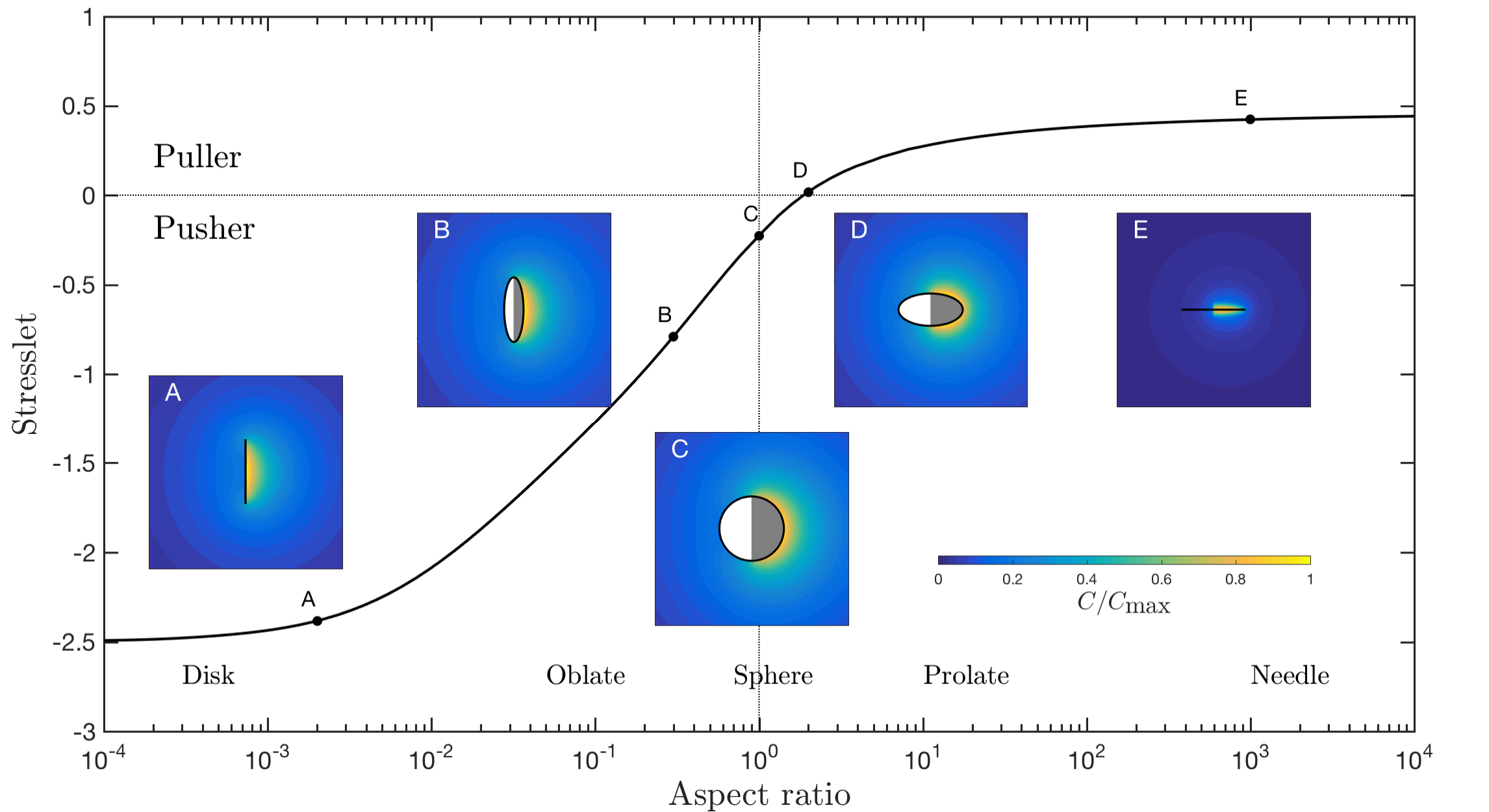}
\caption{Stresslet intensity (scaled by $\mu\mathscr{S}\mathcal{A}_0\mathcal{M}_0/D$) for  a phoretic Janus particle of  spheroidal shape as a function of its aspect ratio with $\mathcal{A}(\zeta)=\mathcal{A}_0$ and $\mathcal{M}(\zeta)=\mathcal{M}_0$ on the active right half (grey), and $\mathcal{A}(\zeta)=\mathcal{M}(\zeta)=0$ on the inert left half (white). The sign of stresslet is reversed by changing the sign of either $\mathcal{A}_0$ or $\mathcal{M}_0$, but not both. The distribution of concentration is also shown.}\label{fig:stresslet}
\end{center}
\end{figure*}

This new result,  impossible  to compute directly analytically otherwise, allows to characterise the role of geometry  on the strength of the stresslet for active particles. For illustration, let us focus on a Janus particle with an active half ($\zeta>0$) of uniform activity and mobility, and an inert half ($\zeta<0$) where both quantities are zero. We plot in Fig.~\ref{fig:stresslet} the strength of the stresslet as a function of the aspect ration of the Janus particle, showing the critical role of geometry. For positive activity (i.e.~solute release on the surface of the particle) and positive mobility (i.e.~slip velocity in the same direction as the local  concentration gradient), oblate particles act as pushers swimmers ($S<0$) while most prolate particles are pullers ($S>0$). The spherical limit ($\xi=1$) corresponds to a weak pusher swimmer while the pusher-puller transition occurs for a blunt prolate with aspect ratio $\xi\approx 2$ (Fig.~\ref{fig:stresslet}). 

These results can be rationalised physically by inspecting the distribution of solute around the particle (see Fig.~\ref{fig:stresslet}, insets). For an oblate or spherical particle, the highest solute concentrations are found at the active pole. The slip velocity along the active boundaries is therefore oriented from the equator to the pole leading to a pusher-type signature on the flow. In contrast for a prolate phoretic particle, the sharp local curvature near the active pole results in a local minimum of the concentration at the pole (chemical solute is efficiently diffused away from that point) and the absolute maximum of the surface concentration is instead found at an intermediate position on the active half of the particle. When $\xi\rightarrow\infty$, one can show that this local maximum of concentration is found at a distance $z_\textrm{max}\approx 0.2 a$ away from the equator. In that case, the slip velocity is still oriented from the equator to the pole for $0\leq z\leq z_\textrm{max}$ but in the reverse direction for $z_\textrm{max}\leq z\leq a$,  the latter being dominant and inducing a puller signature. 

In summary we outlined in this work a new method, based on the  reciprocal theorem for Stokes flows, to compute the stresslet generated by an active particle. The method requires  knowledge of (i) the  instantaneous geometry of the particle, (ii) the prescribed slip velocity along its boundary  and (iii) a dual Stokes problem of an identical rigid particle in an  linear flow. The main advantage of this approach is that it does not require to solve for the actual flow field around the active particle.  After the formal derivation of the method, we  verified it for the classical cases of active spheres and rods for which an alternative,  direct calculation is possible~\cite{SI}. We then  demonstrated how to use our new integral formulation to derive a result impossible to obtain directly, namely the stresslet for spheroidal phoretic particles.

As an extension for future work, we note that when the particle is not torque-free, the present approach could easily be generalized to compute the rotlet generated by the active particle   (i.e.~the strength of the torque locally induced by the swimmer)  by repeating the analysis presented in this paper with a dual flow field where the second-rank tensor $\E$ is  antisymmetric.

We envision our method to be particularly relevant to fixed-shape phoretic swimmers where the dual problem can be solved once and for all. The result of Eq.~\eqref{finalfinal} could then be directly used to sculpt the strength of the stresslet as a function of the chemical and geometrical characteristics of the particle, allowing to potentially tune  interactions of active particles with boundaries and to create active fluids with pre-designed collective or rheological characteristics.

\smallskip
\begin{acknowledgements}
Funding from the EU (CIG to EL) and by the French Ministry of Defense (DGA to SM) is gratefully acknowledged.
\end{acknowledgements}
%\bibliography{stress_refs}
%\end{document}

\end{document}